\documentclass[sigconf]{acmart}

\usepackage{booktabs}
\usepackage{graphicx}
\usepackage{xcolor}
\usepackage{multirow}
\usepackage{enumitem}
\usepackage{amsmath}
\begin{document}

\title{LLM-Driven Intent-Based Privacy-Aware Orchestration Across the Cloud-Edge Continuum}

\author{Zijie Su}
\affiliation{%
  \institution{DisNet Lab, The University of Melbourne}
  \city{Melbourne}
  \state{Victoria}
  \country{Australia}
}
\email{zisu1@student.unimelb.edu.au}

\author{Muhammed Tawfiqul Islam}
\affiliation{%
  \institution{DisNet Lab, The University of Melbourne}
  \city{Melbourne}
  \state{Victoria}
  \country{Australia}
}
\email{tawfiqul.islam@unimelb.edu.au}

\author{Mohammad Goudarzi}
\affiliation{%
  \institution{Monash University}
  \city{Melbourne}
  \state{Victoria}
  \country{Australia}
}
\email{Mohammad.Goudarzi@monash.edu}

\author{Adel N. Toosi}
\affiliation{%
  \institution{DisNet Lab, The University of Melbourne}
  \city{Melbourne}
  \state{Victoria}
  \country{Australia}
}
\email{adel.toosi@unimelb.edu.au}

\begin{abstract}
The rapid growth of edge-to-cloud applications has intensified the need to guarantee that sensitive data remains within trusted jurisdictions while still meeting performance targets. Existing orchestration stacks---Kubernetes for compute and software-defined networking (SDN) controllers such as ONOS for traffic steering---expose rich policy knobs, yet they still require experts to translate high-level privacy goals into dozens of low-level rules.
This paper proposes and experimentally validates an LLM-driven, intent-based framework that enforces privacy requirements across both computing and networking layers of the cloud--edge continuum. A GPT-4o model interprets natural-language privacy intents, then automatically (i) generates Kubernetes node-selector rules to place pods only on nodes that satisfy privacy and locality constraints, and (ii) compiles SDN flow rules that confine packets to compliant paths. To benchmark the approach, we curated a corpus of 90 diverse privacy intents spanning computing, networking, and hybrid scenarios and built an automated validator that integrates a Kubernetes + ONOS test-bed, deploys model-generated configurations, and issues pass--fail reports without human intervention. Experiments show that GPT-4o converts intents into correct, enforcement-ready policies in 95.6\% of 90 end-to-end trials while holding average latency to around 21 seconds. The unified validator further demonstrates that continuous, cross-layer compliance checking can be executed in seconds, eliminating hours of manual inspection and reducing error rates.
\end{abstract}



\maketitle

\section{Introduction}

The rapid increase of devices on the Internet of Things (IoT) has led to an explosion of data being generated at the network edge. Traditionally, centralized cloud computing infrastructures have been used to process and store these data. However, offloading all computation to the cloud is increasingly problematic for applications requiring real-time responsiveness or handling sensitive information. Cloud-only solutions introduce significant network latency and raise privacy concerns regarding the transmission of personal or sensitive data~\cite{akbari2025intentcontinuum}. Additionally, regulatory requirements (e.g., data residency laws) increasingly demand that certain data remain within specific jurisdictions, e.g., the EU General Data Protection Regulation (GDPR)~\cite{eu2016gdpr}. As a result, systems must carefully control placement and routing decisions to ensure compliance. For example, an autonomous vehicle that must react to an obstacle cannot afford the delay in sending sensor data to a distant cloud for processing~\cite{shi}, while regulatory policies may require that sensitive data remain within the legal jurisdiction of the driver’s country. Similarly, a healthcare IoT device that collects patient information may be legally required to keep data within local hospitals or national borders, despite having more relaxed latency requirements.~\cite{iot-medical}.

These pressures have catalysed the emergence of the \textbf{computing continuum}: a heterogeneous fabric of edge and cloud resources that collaborates to meet both performance and compliance needs. Because data and computation are dispersed across this continuum, systems must determine \emph{where} each task and dataset should reside so that latency targets are achieved while data stay in permitted locations. Consequently, research now focuses on \emph{privacy-aware placement}---automated techniques that decide, for every intent, which nodes can host a service and which network paths satisfy jurisdictional constraints~\cite{continuum_intro}. For instance, a user or application might specify an intent such as: ``Ensure all personal data remains within the European Union''---a directive aligned with GDPR data residency requirements. Such natural-language intents must be translated into concrete infrastructure-level configurations, like binding pods to EU-based nodes and restricting network paths to remain within EU jurisdictions.

\textbf{Intent-driven management} has matured into a mainstream paradigm for operating cloud, edge, and network infrastructures~\cite{Sharma}. Intent-based management systems enable developers and infrastructure engineers to specify their intents, such as privacy and latency, allowing the platform to adapt autonomously. In industrial controllers as well as academic prototypes, high-level goals are expressed in a domain-specific language (DSL) or structured policy model and then compiled into concrete configurations~\cite{intent_k8s}. A complementary challenge now emerges: enabling \emph{non-specialists}---for example business analysts or compliance officers---to express intent in ordinary natural language, and letting the platform interpret and enforce those requirements on the fly.

Recent advances in generative AI and large language models (LLMs) create a promising path toward that objective. Models such as GPT-4o~\cite{gpt4} can translate complex instructions into well-formed code or configuration, reason over ambiguous statements, and adapt their output to real-time context~\cite{hagos2024recent,translation_toward}. These findings motivate the systematic investigation presented in this paper: employing an LLM as a natural-language interface to an intent compiler, while integrating with established domain-specific mechanisms to enforce policies.

\begin{figure}[t]
  \centering
  \includegraphics[width=0.95\linewidth]{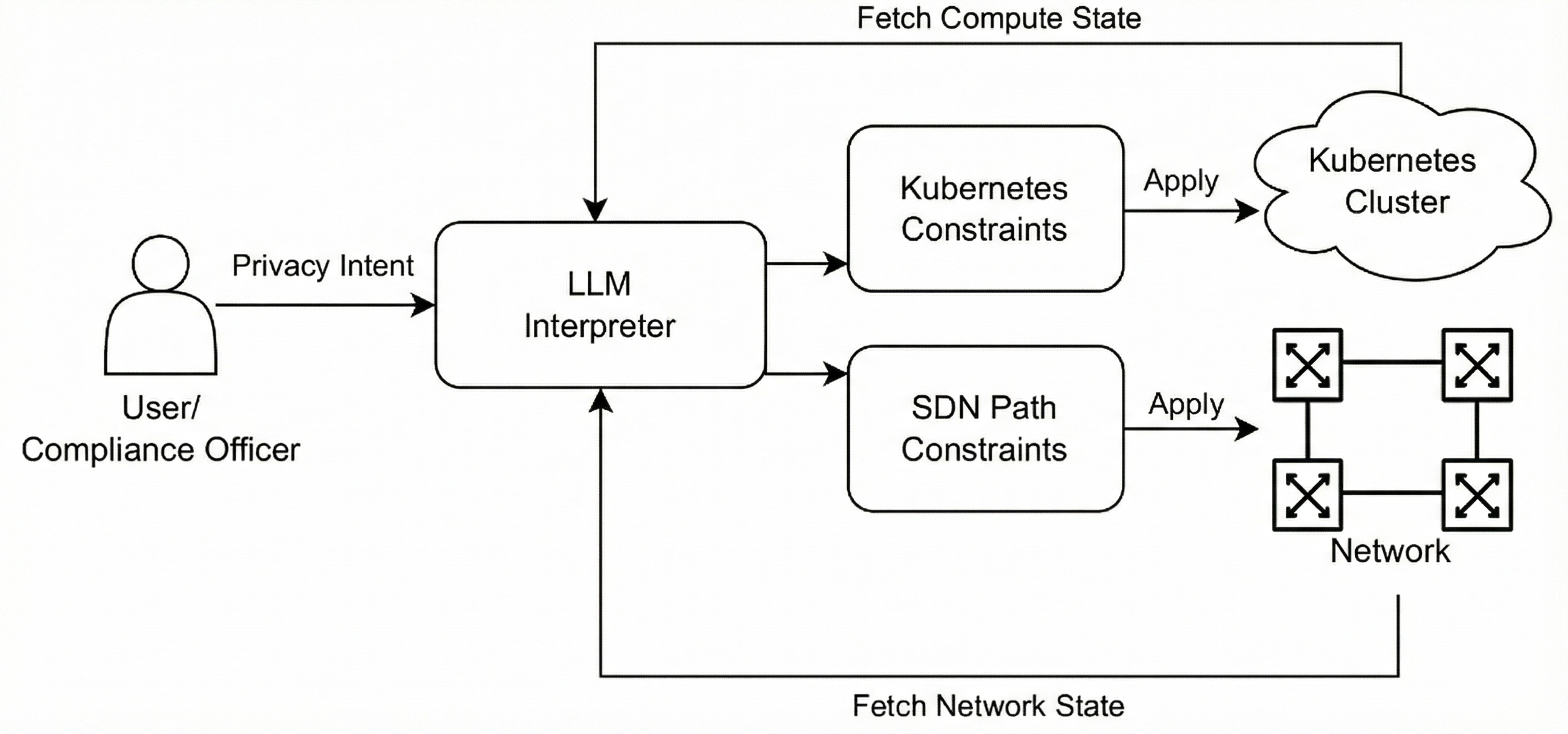}
  \caption{High-level architecture of the LLM-driven, intent-based privacy-preserving orchestration framework.}
  \label{fig:overview}
\end{figure}

Building on these insights, this paper explores an LLM-driven, intent-based approach to privacy-preserving orchestration in the computing continuum. We focus on privacy intents---high-level directives that express user or application preferences regarding data privacy and locality. A LLM is used to interpret these intents and enforce them through the coordinated scheduling of computing and network resources. Unlike previous frameworks that primarily target performance optimization, this work is the first to concentrate specifically on privacy requirements across both cloud-edge environments and network infrastructures. At the computing layer, the system influences Kubernetes pod scheduling by selecting appropriate nodes based on security attributes. At the network layer, it configures the Software-Defined Networking (SDN) controller to establish communication paths that align with the specified privacy requirements. Figure~\ref{fig:overview} provides an architectural overview of our cross-layer intent compilation and enforcement across Kubernetes and an SDN Controller.

Despite growing interest in intent-driven orchestration, the community still lacks a standard benchmark to evaluate how accurately different methods translate natural-language privacy directives into enforceable policies. Existing studies~\cite{klonetai2025} often rely on limited, ad-hoc test cases or manual verification. To address this gap, we construct a benchmark dataset comprising 90 natural-language privacy-related intents that span a diverse set of enforcement scenarios, including data locality, infrastructure avoidance, trust zone constraints, and provider restrictions. In parallel, we implement an automated validator that works with a hybrid Kubernetes and ONOS test-bed, applies model-generated configurations, captures the resulting run-time state, and produces pass--fail compliance reports.

The key \textbf{contributions} of this paper are as follows:
\begin{itemize}[leftmargin=*]
  \item \textbf{Intent-Driven Privacy Framework:} We design and implement a framework that integrates a LLM into the orchestration loop, enabling interpretation of natural-language privacy intents and automated enforcement across a cloud--edge continuum, with coordinated \emph{compute + network} orchestration.
        
  \item \textbf{Coordinated Compute and Network Scheduling:} Our system extends intent-based resource management to jointly handle Kubernetes-based pod scheduling \emph{and} SDN path configuration, ensuring that sensitive data not only resides on appropriate nodes but also travels through compliant network routes.
        
  \item \textbf{Prototype Implementation and Verification:} We present a prototype using Kubernetes for container orchestration, ONOS controlling a network, and GPT-4o for intent interpretation. Privacy intents are translated into Kubernetes directives and ONOS flow rules.
        
  \item \textbf{Benchmark Dataset and Automated Validator:} We curate a dataset of 90 privacy intents and provide a one-click validator that integrates a Kubernetes + ONOS test-bed, deploys generated policies, and checks run-time compliance.
\end{itemize}

 The remainder of this paper is organized as follows. Section~2 reviews related work on privacy-aware scheduling, intent-based management, and LLM-assisted policy generation, positioning our contributions. Section~3 formalizes the problem, including assumptions, threat model, and a satisfaction-based view of intent compliance. Section~4 presents the end-to-end system design, explaining how we translate intents into Kubernetes placement constraints and SDN path constraints. Section~5 details the experimental setup, including the Kubernetes+ONOS testbed, the workload and label schema, and the benchmark intent corpus. Section~6 reports end-to-end accuracy/latency/cost results and analyzes failure modes and practical implications. Section~7 concludes and outlines future work.

\section{Related Work}



Modern edge-cloud schedulers address privacy through data-centric protections and compute-side placement policies. Early approaches focused on transforming data to reduce sensitivity before offloading from resource-constrained devices. Data aggregation and noise injection techniques obscure individual details by combining fine-grained IoT streams into coarser statistics~\cite{aggregation_privacy}. For latency-sensitive AI applications, splitting deep neural networks so that sensitive early layers remain on-device while deeper layers execute on edge servers~\cite{shi}.

Sharif et al.~\cite{Sharif2017PrivacyAware} address privacy in hybrid clouds by designing scheduling algorithms that keep sensitive tasks on private resources. Their Privacy-Aware Scheduling broker balances execution cost against privacy constraints. Wang et al.~\cite{wang_scheduling} formulate scheduling of workflows containing tasks with multiple privacy levels across a cloud--edge--end architecture, proposing a heuristic scheduler (PWHSA) that classifies each task's privacy sensitivity. Alonso-Lupez et al.~\cite{level_of_trust} use reinforcement learning to optimize service orchestration while evaluating trust levels of nodes.


Intent-based networking (IBN) allows network administrators to declare what the network should achieve rather than specifying how~\cite{Sharma}. Ujcich and Sanders embed GDPR concepts into ONOS: flows are annotated with fields such as purpose and lawful-basis, while switches are tagged with jurisdiction and encryption capability~\cite{ujcich2019data}. Budhraja et al.~\cite{Budhraja} assign routes privacy-exposure and compliance risk, then invoke heuristics to select low-risk paths. Karmakar et al.~\cite{Karmakar2023} equip OpenFlow switches with TPM for attestation reports converted into real-time trust scores.

Tu et al.~\cite{translation_toward} use LLMs to translate network configuration intents into JSON, finding great potential but requiring expert verification. Dzeparoska et al.~\cite{policy_generation} explore mapping users' network intent to policies through prompt engineering and few-shot learning. Fuad et al.~\cite{FUAD2024} introduced an intent-based network management framework integrating GPT with RAG for automating network configuration.


Barrachina-Muñoz et al.~\cite{intent_k8s} present an intent-based 5G orchestrator using Kubernetes API for dynamic application migration. Filinis et al.~\cite{Intent_driven_orchestration} introduced a framework for serverless computing across the continuum with machine learning to optimize placement and scaling. Parra-Ullauri et al.~\cite{kubeflower} propose kubeFlower, a Kubernetes operator that leverages pod affinity/anti-affinity and differential privacy for federated learning clients.


Zafeiropoulos et al.~\cite{Zafeiropoulos} proposed an intent-driven method to orchestrate compute and network resources using a Meta-orchestrator. He et al.~\cite{vehicle} proposed an intent-based framework for vehicular edge computing, addressing both compute and network management. Morichetta et al.~\cite{3_layer} developed a three-layer intent-driven management system managing scaling and placement of serverless functions based on SLOs.


Table~\ref{tab:intent_privacy_studies} summarizes existing work. Three key gaps emerge: (1) LLMs remain underexplored for intent-based management, (2) limited studies are addressing \emph{both} networking and computing resources jointly, and (3) it remains challenging to implement privacy-preserving technologies that are both economical and effective. Our work addresses these gaps by using an LLM to translate natural-language privacy intents into coordinated Kubernetes and ONOS configurations.

\begin{table}[t]
\centering
\scriptsize
\caption{Comparison with related work on intent-based scheduling and privacy control.}
\label{tab:intent_privacy_studies}
\begin{tabular}{p{2cm}p{1.8cm}p{1.4cm}p{1.9cm}}
\toprule
\textbf{Work} & \textbf{Technique} & \textbf{Managed Layers} & \textbf{Privacy Mechanism} \\
\midrule
Sharif et al.~\cite{Sharif2017PrivacyAware} & MPHC policies & compute (cloud) & Private-cloud placement \\
Wang et al.~\cite{wang_scheduling} & PWHSA heuristic & compute (cont.) & Tiered placement \\
Ujcich and Sanders~\cite{ujcich2019data} & ONOS GDPR tags & network & EU-only links \\
Karmakar et al.~\cite{Karmakar2023} & TPM attestation & network & Trust-score routing \\
Van Tu et al.~\cite{translation_toward} & LLM translation & network & --- \\
Fuad et al.~\cite{FUAD2024} & GPT + RAG & network & IP/port anonymisation \\
Parra-Ullauri et al.~\cite{kubeflower} & Kubeflower operator & compute (cloud-edge) & Isolation + diff.\ privacy \\
Morichetta et al.~\cite{3_layer} & Three-layer intent & compute+net & --- \\
\midrule
\textbf{This Work} & LLM parser + generator & \textbf{compute+net} & \textbf{Privacy-compliant placement \& routing} \\
\bottomrule
\end{tabular}
\end{table}

\section{Problem Formulation}

Contemporary cloud-edge infrastructures lack a straightforward mechanism for users to enforce high-level privacy requirements on the placement and flow of their data. Existing solutions either require manual configuration of complex policies through low-level YAML files for Kubernetes and flow rules for SDN, or they are limited to specific, pre-defined categories of data sensitivity.

\textbf{Research Aim:} \textit{To develop a software system which enables intent-driven privacy preservation in a computing continuum, allowing users to specify privacy requirements in natural language and automatically orchestrating both compute and network resources to meet those requirements.}

\subsection{Assumptions and Threat Model}

We assume a \emph{trusted orchestration core} (Kubernetes control plane and SDN controller) that can read infrastructure state (labels and topology) and apply policies. We further assume enforcement labels (e.g., \texttt{region=EU} and \texttt{trusted=yes}) are provisioned by operators or attestation and are not mutable by application pods.

In practice, these assumptions can be supported by standard control-plane hardening and policy mechanisms: restrict who can modify labels and manifests using Kubernetes RBAC and admission control (e.g., validating/OPA Gatekeeper policies), isolate and secure the SDN controller API, and bind sensitive labels to trusted provisioning (e.g., node bootstrap configuration, inventory services, or attestation-backed metadata). If the control-plane is compromised or labels are inconsistent, our guarantees degrade accordingly; we therefore treat label integrity and controller trust as prerequisites for policy-compliance enforcement.

\textbf{Adversary model:} We consider a network and infrastructure adversary who can (i) observe traffic on non-compliant paths, (ii) attempt to induce misplacement by relying on ambiguous intents, and (iii) exploit configuration mistakes. We do \emph{not} attempt to protect against a fully compromised control-plane (e.g., a malicious Kubernetes scheduler or SDN controller), nor do we provide cryptographic confidentiality if the deployment lacks encryption primitives. Our goal is \emph{policy compliance}: ensure that the deployed placement and routing decisions satisfy the user intent with respect to the trusted labels and observed run-time state.

\subsection{System Model}

We model the problem as follows:
\begin{itemize}[leftmargin=*]
    \item A set of application components (microservice or pods) $P = \{p_1, p_2, \ldots, p_n\}$ that need to be deployed. Each $p_i$ may handle certain data, such as user information or sensor readings, characterized by a privacy level $L(p_i)$ derived from an intent description.
    
    \item A set of nodes (infrastructure) $N = \{n_1, n_2, \ldots, n_m\}$, each with attributes such as location, ownership, and security level. A node $n_j$ might be labeled as \textit{edge}, \textit{region=EU}, \textit{trusted=yes}, etc.
    
    \item A network graph connecting these nodes, managed by an SDN controller. Each network device exposes metadata including attributes like \textit{mfr=HUAWEI}, \textit{protocol=OF\_13}, or \textit{role=MASTER}.
\end{itemize}

\subsection{Formal Model and Satisfaction}

Let $P$ be the set of pods (application components), $N$ the set of compute nodes, and $G=(V,E)$ the network graph whose vertices $V$ include nodes and switches. Let $\mathcal{A}$ be the universe of infrastructure attributes (labels/metadata).

\textbf{Label functions:} We model infrastructure state via label functions $\lambda_N : N \rightarrow 2^{\mathcal{A}}$ and $\lambda_V : V \rightarrow 2^{\mathcal{A}}$, returning the set of attributes attached to each node or network vertex (e.g., $\texttt{region=EU} \in \lambda_N(n)$).

\textbf{Configuration:} A deployed configuration is $C=\langle \sigma, \rho \rangle$, where $\sigma: P \rightarrow N$ assigns each pod to a node (placement), and $\rho$ is a set of routing constraints (e.g., required waypoints, forbidden vertices/edges) that the SDN controller compiles into flow rules.

\textbf{Constraints and satisfaction:} We represent a compiled intent as a set of compute constraints $\Phi_C$ and network constraints $\Phi_N$. We say $C$ \emph{satisfies} intent $I$ under state $\lambda$ (written $C \models_{\lambda} I$) if (i) all placement constraints in $\Phi_C$ hold for $\sigma$ given $\lambda_N$, and (ii) all routing constraints in $\Phi_N$ hold for the realized paths induced by $\rho$ in $G$ given $\lambda_V$.

\textbf{Example (intent $\rightarrow$ constraints):} Consider the intent: \emph{``Place PHI-processing pods only on EU nodes, and ensure their traffic avoids untrusted switches.''} The compiler produces (a) a placement constraint $\forall p \in P_{\textsf{PHI}}:~\texttt{region=EU} \in \lambda_N(\sigma(p))$, and (b) a routing constraint that all paths carrying PHI flows must exclude vertices labeled \texttt{trusted=no}, i.e., $\forall v$ on the path: $\texttt{trusted=no} \notin \lambda_V(v)$. In Kubernetes, (a) becomes a node selector/affinity on \texttt{region=EU}; in SDN, (b) becomes a forbidden-vertex policy compiled into flow rules.

\textbf{Problem Definition:} Given a privacy intent $I$ expressed by the user in natural language, determine:
\begin{enumerate}[leftmargin=*]
    \item A mapping of each application component $p_i \in P$ to a node $n_j \in N$, such that privacy requirements of $I$ are satisfied with respect to data placement.
    
    \item A set of network routing policies such that any communication involving those components is carried over paths compliant with $I$.
    
    \item Among all solutions satisfying $I$, select one that optimizes secondary objectives such as load balancing and/or performance, without violating the privacy intent.
\end{enumerate}

This problem combines scheduling (NP-hard allocation) with global privacy constraints. Privacy zones may span multiple nodes and network paths. Furthermore, user intents might be incomplete or require interpretation---unlike strict mathematical constraints, intents such as ``use secure infrastructure'' must be translated into specific, actionable criteria.

\subsection{LLM-Driven Intent Interpretation}

In our system, privacy intents are entered through a command-line interface accepting free-form natural language. The CLI sends this input to the intent interpretation module, which invokes a LLM to translate the input into structured policy directives for both compute and network domains.
 
The LLM's contributions can be summarized in two core functions:
\begin{enumerate}[leftmargin=*]
\item \textbf{Semantic Parsing of Ambiguous Expressions:} The LLM interprets free-form, possibly vague, user instructions and translates them into a precise, machine-understandable format. Privacy intents often appear in colloquial terms that do not directly map to system operations---for example, ``our most sensitive data should never leave the cloud.''

\item \textbf{Ontological Linking:} The LLM bridges high-level privacy concepts and the system's low-level identifiers. It can be inferred that ``most sensitive health data'' pertains to Protected Health Information (PHI), relating to pods labeled \texttt{data-type=phi}. By leveraging its training on technical documentation and policy language, the LLM outputs structured intents incorporating correct labels and selectors.
\end{enumerate}

\section{System Design}

This section describes the end-to-end architecture that turns a natural-language privacy intent into \emph{enforcement-ready} compute and network configurations, and explains how the system integrates live infrastructure state to keep decisions consistent with deployment reality.

\begin{figure*}[t]
    \centering
    \includegraphics[width=0.85\textwidth]{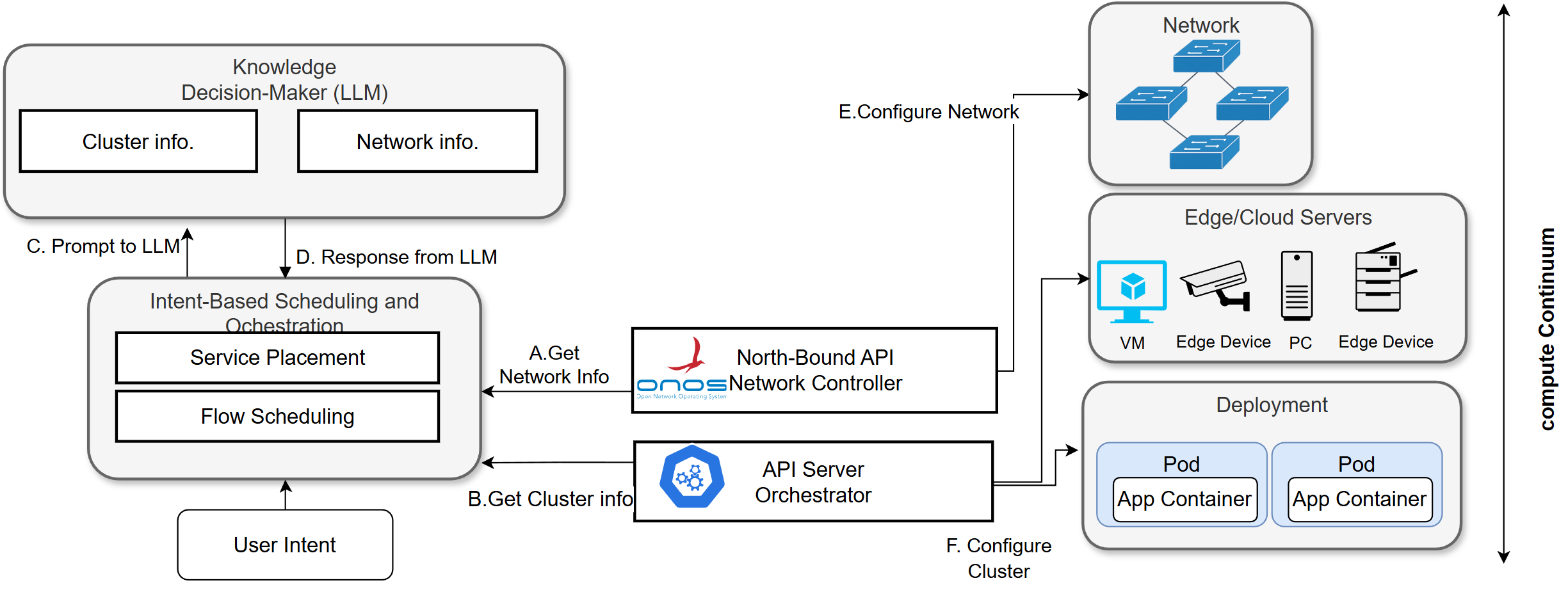}
    \caption{System architecture for intent-driven privacy-aware orchestration.}
    \label{fig:system_architecture}
\end{figure*}

Figure~\ref{fig:system_architecture} provides a high-level view of the control loop and its major components. We organize the design into logical planes: an LLM-driven knowledge plane that produces structured directives, an orchestration plane that compiles and applies those directives to Kubernetes and ONOS, and an infrastructure plane that exposes run-time state for closed-loop validation and adaptation.

\subsection{LLM-reasoning-based Knowledge Plane}

The Knowledge Plane centralizes AI-driven reasoning, leveraging GPT-4o via the OpenAI API. The LLM receives a structured prompt composed of the user intent and a condensed infrastructure snapshot (node labels, pod distributions, network topology, and link status). It returns machine-consumable directives:

\begin{itemize}[leftmargin=*]
\item \textbf{Placement Directives:} Kubernetes node selection criteria and affinity/anti-affinity rules for data locality, regulatory compliance, or security zoning.
\item \textbf{Flow Directives:} Network routing constraints, detailing explicit path selections or device exclusions for vendor avoidance or trusted-path compliance.
\end{itemize}

Isolating AI-based reasoning in a dedicated plane improves auditability and enables the downstream execution stages to remain deterministic.

\textbf{LLM technique design:} Prompting is modular: each sub-task is governed by a role-specific prompt enforcing output formats. Key roles include: (1) an \textbf{Intent Classifier} (compute/network/hybrid), (2) a \textbf{State Checker} to decide which state to retrieve, (3) a \textbf{Service Scheduler} to map constraints into Kubernetes affinities/tolerations, and (4) a \textbf{Path Planner} to emit ONOS-compatible path constraints. Few-shot examples (2--3 pairs) and explicit schemas (``do not include fields outside the specified schema'') reduce hallucinations.

\subsection{Intent-driven Orchestration Plane}

This intermediary plane operationalizes the LLM-derived configurations:

\begin{itemize}[leftmargin=*]
\item \textbf{Service Placement Module:} Translates placement directives into Kubernetes manifests, incorporating affinity rules and scheduling constraints.
\item \textbf{Flow Scheduling Module:} Converts network flow directives into ONOS intents or flow-modification messages via REST APIs.
\end{itemize}

The modules follow a six-step interaction loop: (A) Query ONOS for network topology, (B) Query Kubernetes API for node labels and pod locations, (C) Construct an enriched LLM prompt, (D) Parse LLM response into structured instructions, (E) Execute network flow instructions through ONOS, (F) Apply service placement through Kubernetes.

\textbf{Compute orchestration (service placement):} Given a deployment intent, the system produces a complete Kubernetes manifest. Prompts are crafted so LLM yields multi-document YAML blocks along with shell commands like \texttt{kubectl apply\allowbreak -f\allowbreak <manifest>.yaml}, enabling one-step deployment.

\begin{figure*}[t]
  \centering
  \includegraphics[width=0.8\textwidth]{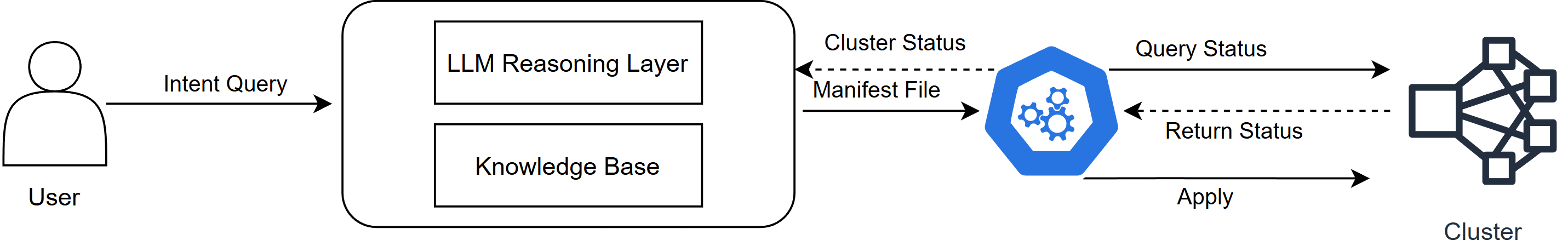}
  \caption{LLM-assisted service placement workflow.}
  \label{fig:llm-k8s-architecture}
\end{figure*}

Figure~\ref{fig:llm-k8s-architecture} shows the high-level loop for compute intents: the user provides a natural-language intent, the knowledge plane (LLM reasoning + knowledge base) interprets it into placement constraints, emits Kubernetes manifests, and applies them to the cluster.

\begin{figure*}[t]
  \centering
  \includegraphics[width=0.8\textwidth]{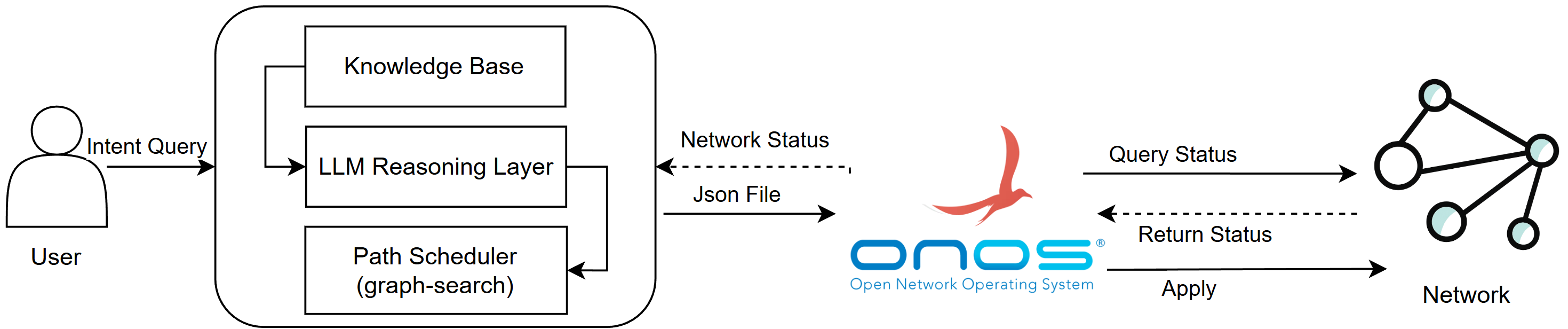}
  \caption{LLM-assisted network path control workflow.}
  \label{fig:llm-onos-routing}
\end{figure*}

\textbf{Network orchestration (data-flow control):} The system translates high-level user intents into enforcement-ready flow configurations. Figure~\ref{fig:llm-onos-routing} summarizes the routing loop: the user submits an intent, the knowledge base and LLM reasoning layer translate it into structured constraints, and a path scheduler performs graph search over the current topology to produce a validated path. The resulting JSON intent is then applied via ONOS.

\begin{figure}[t]
  \centering
  \begin{minipage}{0.95\columnwidth}
    \hrule
    \footnotesize
  \begin{verbatim}
[{
  "src": "of:000000000000000X",
  "dest": "of:000000000000000Y",
  "out_port": M,
  "forbidden": ["of:000000000000ZZZ"],
  "must_go": ["of:000000000000AAA"]
}]
  \end{verbatim}
    \hrule
  \end{minipage}
  \caption{Example LLM-generated JSON constraints for ONOS.}
  \label{lst:json-constraints}
\end{figure}

This structured output allows the system to apply graph algorithms (weighted Dijkstra, BFS fallback) to compute valid paths respecting constraints. Each path is transformed into per-hop flow rules submitted to ONOS via REST API.

\textbf{Hybrid coordination:} Compute-layer decisions must precede network-layer enforcement because endpoints become concrete only after pods are scheduled. The pipeline therefore proceeds: first allocate pods, then observe resulting attachments/topology, then compile and install flow rules restricting which links and switches each flow may traverse.
\subsection{Infrastructure Plane with State Integration}

The Infrastructure Plane constitutes the concrete execution layer:

\begin{itemize}[leftmargin=*]
    \item \textbf{ONOS SDN Controller:} Exposes a north-bound intent API for installing flow rules and collecting real-time telemetry. All network-level privacy constraints are enforced here.

    \item \textbf{Kubernetes API Server:} Acts as the authoritative control plane for pod scheduling, maintaining node labels (\texttt{region}, \texttt{trust\_zone}, \texttt{vendor\_avoid}).

    \item \textbf{Edge/Cloud Nodes:} Physical and virtual machines running containerised micro-services.
\end{itemize}

\textbf{State collection and integration:} The system is state-aware: it does not enforce intents in a vacuum, but in the context of current cluster and network status. Based on the user's request, the system retrieves relevant state (Kubernetes: node labels, pod placement, resource definitions; ONOS: topology, link state, host attachments, existing flows). The collected data are condensed into a JSON-like dictionary injected into the LLM prompt.

\subsection{Safety Control and Validation}

The system treats LLM output as a \textit{suggested} plan rather than an authoritative directive. A validation and governance layer checks recommendations against domain constraints and operational policies before applying any configuration. If the LLM proposes scheduling a pod on a node lacking necessary labels or suggests a network modification breaching a security rule, the system detects this and either discards or modifies the directive.

\section{Experimental Setup}

All experiments are conducted on an edge-cloud testbed provisioned via the iContinuum emulation toolkit~\cite{iCOntinuum_paper}, deployed on virtual resources from the University private cloud environment.

\subsection{Cluster Configuration}

The experimental platform employs K3s, a lightweight Kubernetes distribution, deployed across a multi-node cluster. Nodes are annotated with metadata (e.g., \texttt{location=region-a}, \texttt{zone=cloud}, \allowbreak\texttt{security=high}) to emulate diverse trust zones. Two configurations were deployed: a 5-worker cluster (9 OpenFlow switches, 30 links) and a 13-worker cluster (25 switches, 74 links).

\textbf{Rationale for two cluster sizes:} The 5-worker cluster provides a compact topology for fast iteration and deterministic debugging of placement and routing constraints. The 13-worker cluster increases infrastructure heterogeneity and the path-search space (more switches/links and more feasible placements), allowing us to evaluate scalability and intent compilation and validation under a larger set of admissible schedules and routes.

\begin{table}[t]
  \centering
  \caption{Virtual-machine specifications for experimental clusters.}
  \label{tab:vm-specs}
  \begin{tabular}{lccc}
    \toprule
    \textbf{Node role} & \textbf{vCPUs} & \textbf{RAM} & \textbf{Count}\\
    \midrule
    Worker       & 2 & 8\,GB  & 5 / 13\\
    Master       & 4 & 16\,GB & 1\\
    ONOS/Mininet & 8 & 32\,GB & 1\\
    \bottomrule
  \end{tabular}
\end{table}

Network connectivity is orchestrated through ONOS, with the underlying topology emulated using Mininet and Open vSwitch. ONOS exposes REST APIs for dynamic control, enabling enforcement of network-level intents such as geographic routing restrictions and vendor exclusions.

\subsection{Application Workload}

To emulate a high-privacy deployment, we modeled a representative hospital information-system workload comprising six micro-services: \textit{Appointment}, \textit{Doctor}, \textit{Patient}, \textit{Vital-Sign Monitor}, \textit{PHI-DB}, and \textit{General-DB}. Services handling protected health information (PHI) require placement on high-trust nodes, while less sensitive services may run on any available node.

\begin{table}[t]
  \centering
  \caption{Microservice workload used in the experiments.}
  \label{tab:microservices_privacy}
  \scriptsize
  \begin{tabular}{p{2.3cm}p{2.0cm}p{2.8cm}}
    \toprule
    \textbf{Service (app)} & \textbf{Sensitivity label(s)} & \textbf{Role} \\
    \midrule
    \texttt{db} & \texttt{data-type} $\in$ \{\texttt{phi}, \texttt{general}\} & Database service with PHI/non-PHI variants \\
    \texttt{phi-db} & \texttt{data-type=phi} & PHI-focused database deployment used by PHI-residency intents \\
    \texttt{patient} & PHI & PHI-processing microservice (patient records) \\
    \texttt{appointment} & General & Application-tier microservice (appointments) \\
    \texttt{doctor} & General & Application-tier microservice (doctor information) \\
    \texttt{vital-sign-monitor} / \texttt{image-preprocessor} & PHI / General & Auxiliary microservices used in the larger setup \\
    \addlinespace[0.2em]
    \multicolumn{3}{p{\linewidth}}{\scriptsize \textbf{Note:} Service identity and data sensitivity are encoded via Kubernetes labels (e.g., \texttt{app}, \texttt{data-type}) and referenced by validator predicates and scheduling constraints.}\\
    \bottomrule
  \end{tabular}
\end{table}

\begin{table}[t]
  \centering
  \caption{Label schema used to compile and validate privacy intents (selected examples).}
  \label{tab:label_schema}
  \scriptsize
  \renewcommand{\arraystretch}{1.05}
\begin{tabular}{p{0.7cm}p{2.0cm}p{2.2cm}p{2.4cm}}
    \toprule
    \textbf{Layer} & \textbf{Label / field} & \textbf{Example values} & \textbf{Constraint semantics} \\
    \midrule
    Service & \texttt{app}, \texttt{data-type} & \texttt{db, patient}; \texttt{phi/general} & Select affected workloads and flows (PHI vs.\ general) \\
    Node & \texttt{node\_labels.zone} & \texttt{edge/cloud} & Data locality / keep PHI off edge \\
    Node & \texttt{node\_labels.location} & \texttt{region-a/region-b/\dots} & Residency constraints (region scoping) \\
    Node & \texttt{node\_labels.provider} & \texttt{aws/azure/\dots} & Provider avoidance / trusted provider policies \\
    Node & \texttt{node\_labels.security-level} & \texttt{high/medium/low} & Trust-zone placement (high-trust only) \\
    Network & \texttt{device\_labels.mfr}, \texttt{device\_labels.protocol} & \texttt{huawei}; \texttt{CISCO}\dots & Vendor/protocol constraints for routing \\
    Network & \texttt{device\_labels.location} & \texttt{region-a/region-b/\dots} & Geography-bounded paths / exclusions \\
    \addlinespace[0.2em]
    \multicolumn{4}{p{\linewidth}}{\scriptsize \textbf{Note:} Intents are compiled into Kubernetes selectors/affinities and SDN path constraints over these labels; the validator checks the post-deployment state for compliance.}\\
    \bottomrule
\end{tabular}
\end{table}

\begin{table}[t]
  \centering
  \caption{Worker-node label matrix for 5-worker cluster.}
  \label{tab:labels}
  \begin{tabular}{lcccc}
    \toprule
    \textbf{Node} & \textbf{location} & \textbf{provider} & \textbf{security} & \textbf{zone}\\
    \midrule
    worker-1  & london        & aws            & high   & edge\\
    worker-2  & newyork       & aws            & medium & edge\\
    worker-3  & sanfrancisco  & azure          & medium & cloud\\
    worker-4  & sydney        & azure          & high   & cloud\\
    worker-5  & beijing       & alibaba-cloud  & low    & cloud\\
    \bottomrule
  \end{tabular}
\end{table}

\subsection{Intent Dataset}

We curated a test set of 90 privacy intents, evenly divided across three domains: \textbf{Computing}, \textbf{Networking}, and \textbf{Hybrid} (both Computing and network conditions). Each domain is further classified by complexity:1) \textbf{simple} intents involve a single condition, and 2) complex intents contain multiple conditions or exceptions. The intent corpus covers common privacy requirements in compute placement and network routing, and refined the wording to include realistic variations. Each intent is aligned with our label schema so it can be compiled into concrete Kubernetes and SDN constraints. Table~\ref{tab:privacy_intent_examples} presents illustrative examples of natural-language privacy intents from our benchmark and the corresponding system-enforced constraints.

\textbf{Distribution:} Our benchmark contains 90 intents in total: 30 computing, 30 networking, and 30 hybrid. By complexity, 38 are simple and 52 are complex; hybrid intents are predominantly complex (28/30) because they combine placement and routing clauses.

\begin{table}[t]
  \centering
  \caption{Illustrative privacy intents from our benchmark (simplified examples).}
  \label{tab:privacy_intent_examples}
  \small
\begin{tabular}{p{1.1cm} p{2.2cm} p{4.0cm}}
    \toprule
    \textbf{Intent Type} & \textbf{Natural-language intent} & \textbf{what it enforces} \\
    \midrule
    \textbf{Computing} &
    \emph{``Prohibit financial database service deployment in the cloud zone.''} &
    Do \emph{not} run the (financial) database on cloud-zone nodes. In our testbed this is treated as \textbf{unenforceable} because no workload matches the required label (e.g., \texttt{app=financial-db}); the system fails closed rather than hallucinating resources. \\
    \midrule
    \textbf{Networking} &
    \emph{``Ensure that all traffic from host~2 to host~4 must traverse the backup switch~s8.''} &
    Enforce a waypoint constraint at the network layer: for the specified flow, the realized ONOS path must include switch~s8. The intent passes only if the validator observes the installed flow path traversing the required waypoint. \\
    \midrule
    \textbf{Hybrid} &
    \emph{``Run appointment only on high-security cloud nodes, enforce that all other hosts communicating with host~4 must pass through the backup switch~s8, and prevent sensitive databases from being deployed in the edge zone.''} &
    A multi-clause policy spanning compute and network: (i) schedule appointment pods only onto cloud nodes labelled \emph{high-security}; (ii) forbid PHI database pods on edge nodes; and (iii) route traffic to host~4 through backup switch~s8. The intent passes only if all clauses hold simultaneously. \\
    \bottomrule
\end{tabular}
\end{table}

We define configuration generation accuracy as the fraction of intents for which the LLM produces a wholly correct configuration:
\[
\text{accuracy} = \frac{\text{successful interpretations}}{90} \times 100\%
\]

\subsection{Candidate Large Language Models}

We evaluate three large language models within our approach:
\begin{itemize}[leftmargin=*]
  \item \textbf{GPT-4o (OpenAI)}~\cite{gpt4} — a flagship proprietary model widely recognized for strong reasoning and general-purpose performance.
  \item \textbf{Claude 3.5 Haiku (Anthropic)}~\cite{anthropic_claude35haiku} — a compact model optimized for fast inference (low-latency responses) and reliable instruction following.
  \item \textbf{DeepSeek-V3 (671B)}~\cite{alibabacloud_deepseek} — a large-scale open-source model with strong reasoning capabilities.
\end{itemize}

All models were evaluated on an identical suite of 90 intents using the same prompt scaffold to ensure a fair comparison. For each run, we recorded four metrics: \emph{(i)} end-to-end success (pass/fail), \emph{(ii)} end-to-end completion time (wall-clock), \emph{(iii)} token usage (prompt+completion), and \emph{(iv)} the number of validator checks executed.

\subsection{Automated Validation Pipeline}

\begin{figure*}[t]
    \centering
    \includegraphics[width=0.75\textwidth,height=0.15\textheight,keepaspectratio]{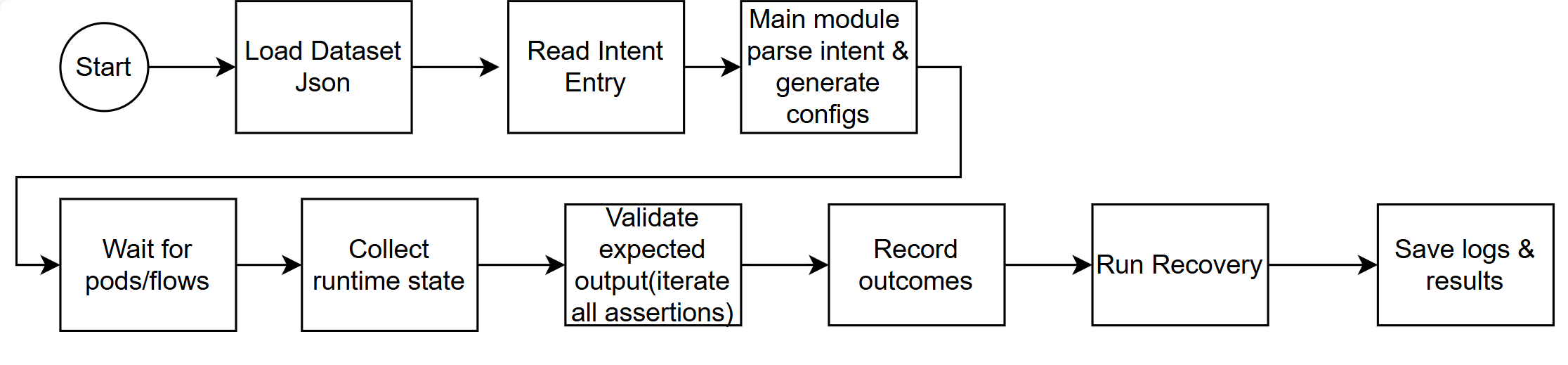}
    \caption{Automated validation pipeline from intent input to result verification.}
    \label{fig:validation-pipeline}
\end{figure*}

To evaluate whether model outputs are \emph{enforcement-ready} (not merely well-formed), we execute an automated validation pipeline that deploys model-generated configurations and checks compliance against the observed run-time state. Figure~\ref{fig:validation-pipeline} summarizes this control loop, and the \texttt{Validator} module orchestrates its execution: loading each experiment, triggering intent interpretation, coordinating platform actions (Kubernetes/ONOS), waiting for pods and flows to stabilize, collecting run-time state, evaluating all assertions, and recording pass/fail outcomes. This structure also makes it explicit which stages contribute to end-to-end latency and where validation is applied.

Each intent entry includes a formal description, validation methods, expected output assertions, and recovery actions.

\textbf{Validator check definition.} A \emph{validator check} is one atomic pass/fail assertion executed by our validator over a post-deployment state snapshot. Each check inspects either the compute state (e.g., a pod is scheduled only onto nodes whose labels satisfy the placement constraints) or the network state (e.g., all realized paths for a target flow avoid forbidden switches / include required waypoints). An intent is marked successful only if all checks for that intent pass; we report the mean number of checks executed per intent as \emph{Avg.\ checks/task}.

\section{Evaluation Results}

\noindent In this section, we evaluate how reliably the proposed framework translates natural-language privacy intents into \emph{enforceable} compute and network policies. We report four primary metrics: (i) end-to-end success rate (pass/fail), (ii) the number of validator checks executed per intent, (iii) end-to-end completion time (wall-clock), and (iv) token usage per query. We first compare candidate LLMs under an identical 90-intent suite to motivate our choice of GPT-4o, then analyze GPT-4o breakdowns by domain and complexity, and finally discuss common failure modes and practical implications.

\subsection{Comparative Evaluation of LLMs}

\begin{figure}[t]
  \centering
  \includegraphics[width=0.92\linewidth]{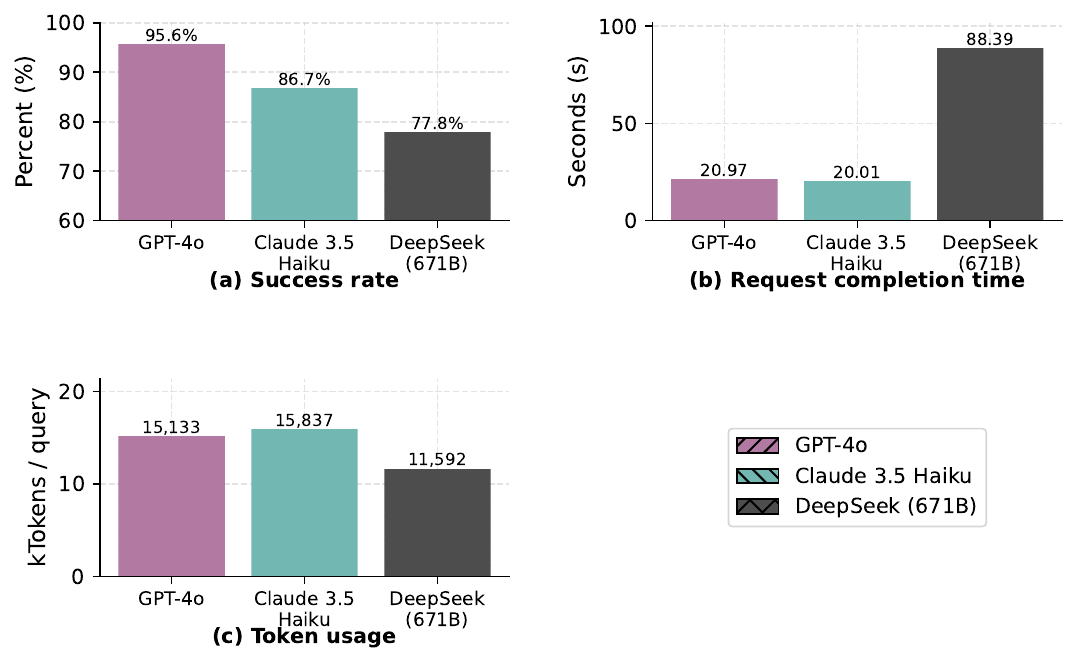}
  \caption{Overall model comparison: (a) success rate, (b) request completion time, and (c) token usage.}
  \label{fig:overall_model_metrics}
\end{figure}

Figure~\ref{fig:overall_model_metrics} contrasts the three models along the core operational axes: success rate, request completion time, and token cost. GPT-4o achieved the highest success rate (95.6\%), significantly outperforming Claude (86.7\%) and DeepSeek (77.8\%). GPT-4o and Claude exhibited comparable average response times of approximately 20 seconds, while DeepSeek was considerably slower at nearly 88 seconds---a result that may be influenced by factors such as API endpoint geography in our setup and model scale/architecture effects (DeepSeek-V3 671B), both of which can affect end-to-end completion time.

\begin{figure}[t]
  \centering
  \includegraphics[width=0.85\linewidth]{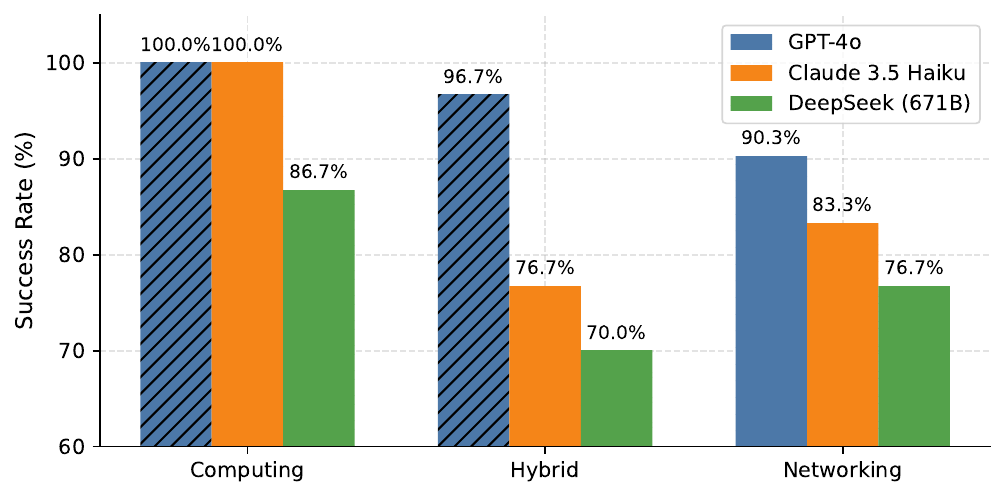}
  \caption{Success rates by domain for each model.}
  \label{fig:domain_success}
\end{figure}

Figure~\ref{fig:domain_success} shows that both GPT-4o and Claude achieved 100\% on computing tasks, while DeepSeek trailed at 86.7\%. On hybrid tasks, GPT-4o retained 96.7\%, while Claude dropped to 76.7\% and DeepSeek to 70.0\%. GPT-4o's success rate varied by only nine percentage points across domains, while Claude's and DeepSeek's dropped by 23.3 and 16.7 points, respectively. Given these results, we use GPT-4o as the primary model for the remainder of our analysis.

\subsection{GPT-4o Performance Analysis}

GPT-4o achieved an overall \textbf{95.6\%} success rate while maintaining an average end-to-end completion time of 20.97 seconds. Importantly, this completion time reflects the \emph{full pipeline} (LLM inference, configuration compilation, Kubernetes/ONOS application, stabilization waits, and post-deployment validation). As a result, time increases are not only ``model latency'' but also signal more expensive orchestration actions (e.g., retrieving additional state or installing more flows).

\begin{table}[t]
  \centering
  \caption{GPT-4o overall performance. \emph{Avg.\ checks/task} counts the number of atomic validator pass/fail assertions evaluated over the post-deployment state for each intent.}
  \label{tab:gpt4o_overall}
  \begin{tabular}{lcc}
    \toprule
    \textbf{Metric} & \textbf{Value} & \textbf{Notes} \\
    \midrule
    Tasks analysed                & 90            & --- \\
    Overall Accuracy              & 95.6\%        & --- \\
    Avg.\ checks / task           & 3.7           & Validator assertions \\
    Avg.\ completion time / task (s) & 20.97      & Wall-clock time \\
    Avg.\ tokens / task           & 15,133        & Prompt + completion \\
    \bottomrule
  \end{tabular}
\end{table}

\noindent \textbf{Interpreting validator checks.} The reported \emph{Avg.\ checks/task} counts the number of atomic validator assertions evaluated over the post-deployment state, not the number of times the LLM is re-invoked. In practice, a single multi-clause intent can trigger several independent checks (e.g., placement and routing clauses), and each clause may require multiple state-retrieval calls before a pass/fail decision is recorded.

\begin{figure}[t]
  \centering
  \includegraphics[width=0.92\linewidth]{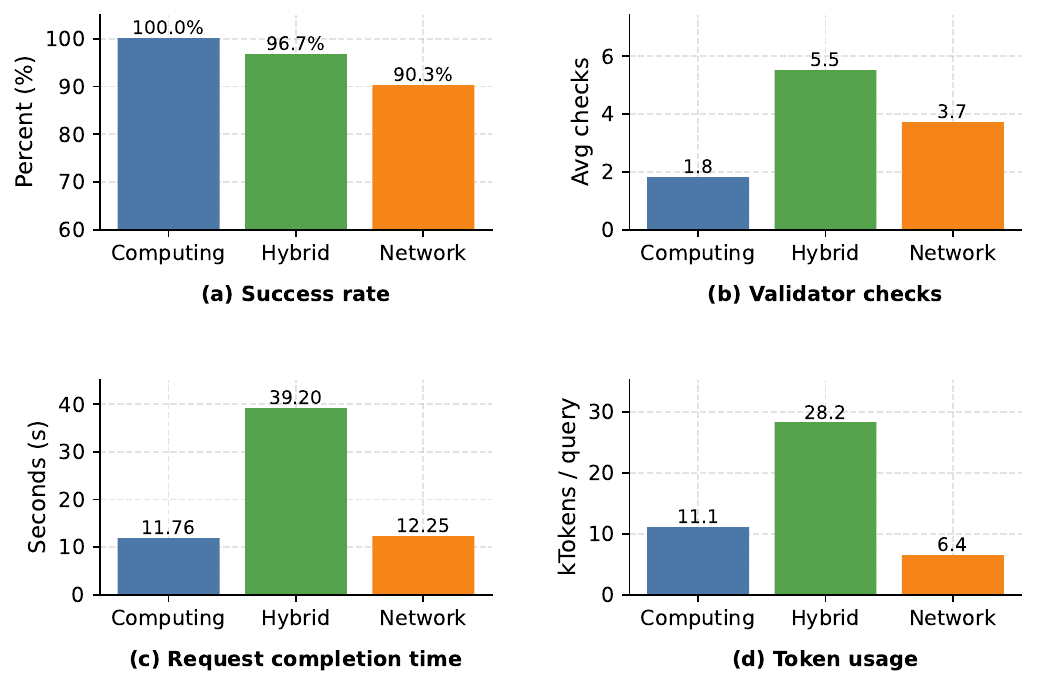}
  \caption{GPT-4o performance by domain: (a) success rate, (b) validator checks, (c) request completion time, and (d) token usage.}
  \label{fig:gpt4o_domain_metrics}
\end{figure}

\textbf{Domain-specific performance.} Figure~\ref{fig:gpt4o_domain_metrics} provides a multi-metric view of domain behavior. The success-rate panel shows accuracy remains near ceiling for computing (100.0\%) and hybrid (96.7\%) intents, while networking intents are more error-prone (90.3\%). The validator-checks panel indicates that hybrid intents trigger substantially more assertions (5.5 checks on average) than computing (1.8) and networking (3.7), reflecting that hybrid sentences often contain multiple independent clauses spanning compute and network constraints. The completion-time and token-usage panels show the corresponding cost of this coupling: hybrid intents take 39.20\,s on average (vs.\ 11.76\,s computing and 12.25\,s networking) and consume 28,207 tokens per query (vs.\ 11,083 computing and 6,399 networking), due to additional state context and cross-layer directives.

Compute-side constraints are grounded in a relatively stable label space (node labels and pod selectors) and are enforced by a deterministic scheduler. In contrast, network intents require grounding \emph{endpoints} and \emph{paths} in a topology that can change with placement decisions, and the enforcement surface is more brittle: a partially specified constraint can silently compile into a no-op policy (no applicable flow), which the validator correctly flags as a failure. This explains why networking has lower success even when its mean completion time is close to computing: the effort is not purely ``more work'' but ``more opportunities to mis-specify.''

\begin{figure}[t]
    \centering
    \includegraphics[width=0.8\linewidth]{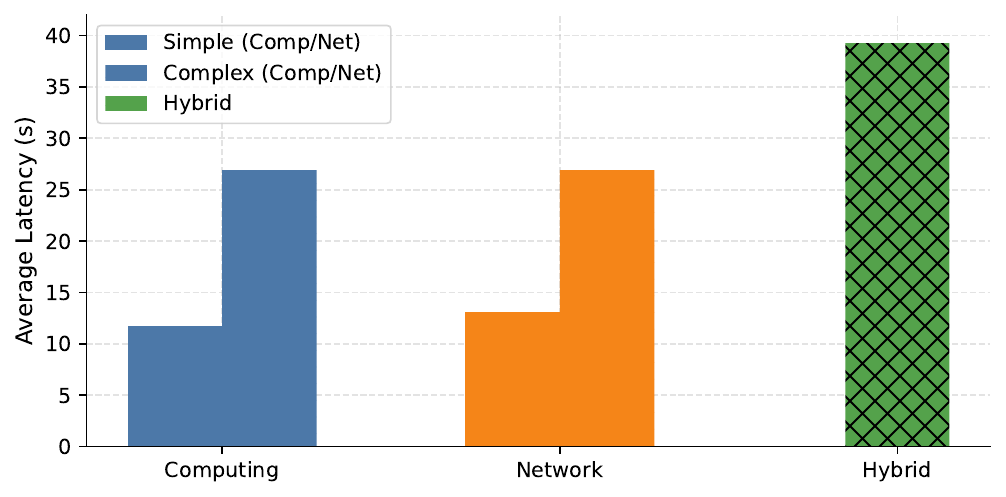}
    \caption{Average end-to-end completion time by intent type and complexity for GPT-4o.}
    \label{fig:latency}
\end{figure}

Figure~\ref{fig:latency} illustrates end-to-end time costs. In the Computing group, simple requests finish fastest (11s) while complex requests take more than double that time (25s). Hybrid intents incur the highest delay at 39s. Despite these differences, accuracy stays high overall: 95\% on computing and hybrid workloads, dipping only slightly (88\%) on network-only tasks.

\begin{figure}[t]
  \centering
  \includegraphics[width=0.92\linewidth]{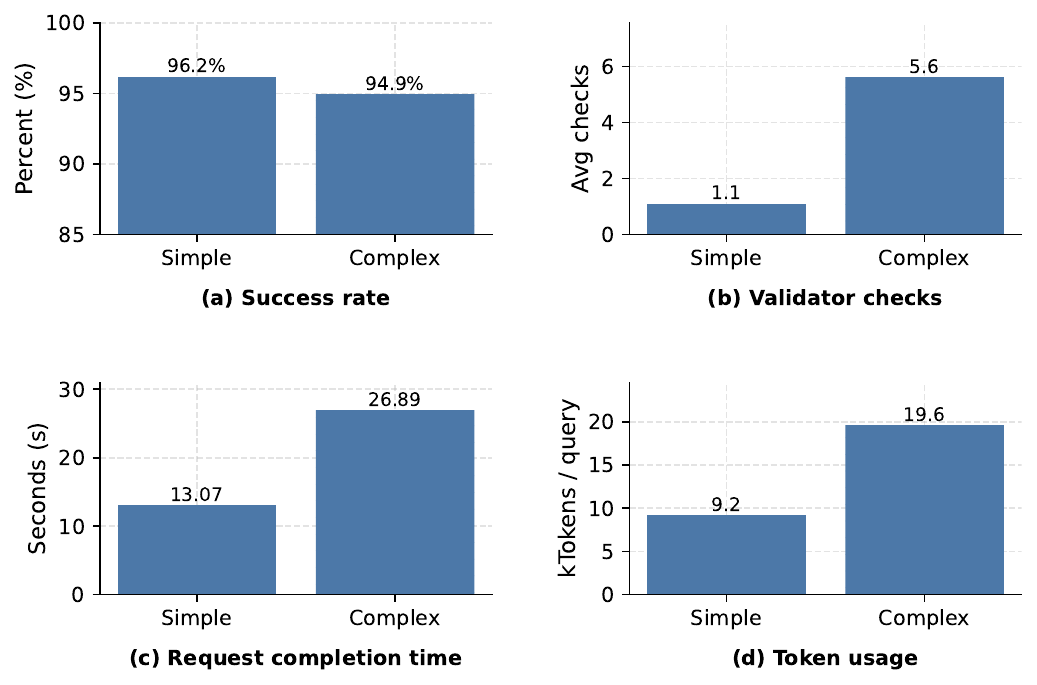}
  \caption{GPT-4o performance by task complexity: (a) success rate, (b) validator checks, (c) request completion time, and (d) token usage.}
  \label{fig:gpt4o_complexity_metrics}
\end{figure}

\textbf{Effects of intent complexity.} Figure~\ref{fig:gpt4o_complexity_metrics} shows that increased intent complexity primarily raises validation effort and end-to-end cost: complex intents require 5.6 checks on average (vs.\ 1.1 for simple) and take 26.89\,s on average (vs.\ 13.07\,s for simple), consistent with multi-clause policies that trigger more independent assertions. The 1.3-point gap between complex and simple success rates lies within the noise band for our study ($\pm$3\% Wilson interval).

\noindent \textbf{Why the simple-vs-complex gap is inconclusive.} With a 90-intent corpus, a handful of outcomes can shift the measured success rate by multiple percentage points. Moreover, the ``simple'' group contains terser prompts that can be ambiguous (leading to parsing errors), while complex intents sometimes include overlapping constraints that make it easier for a partial solution to still satisfy the validator. A larger, less ambiguous corpus would better isolate the effect of complexity on accuracy.

\textbf{Practical implications.} Hybrid intents are consistently the hardest in our setting because they couple placement decisions with topology-dependent routing constraints, increasing both the number of validator checks and the end-to-end latency. In practice, this suggests deployments should prefer (i) lightweight intent decomposition (split compute vs.\ network clauses), and (ii) schema-constrained prompting for network directives, so that operators obtain predictable enforcement-ready outputs with fewer corrective iterations. More broadly, the \emph{validator-check count} provides a useful early warning signal: intents that trigger many checks tend to correspond to multi-clause policies, and are therefore good candidates for either decomposition or more explicit constraint enumeration.

\subsection{Common Failure Modes}

Post-hoc inspection revealed four systematic error sources:
\begin{enumerate}[leftmargin=*]
  \item \textbf{Semantic congestion in hybrid prompts:} When placement, routing, and data-type clauses are fused into a single sentence, the LLM often satisfies only the first clause encountered (``first-clause capture'').

  \item \textbf{Ambiguous path specification:} The SDN sub-pipeline expects explicit \texttt{<src, dst, must\_go>} triples. Prose like ``hosts communicating with host~4 must traverse the backup switch'' lacks concrete pairs, causing the parser to emit an empty object.
  
  \begin{quote}\footnotesize
  \emph{Example:} The LLM returns a constraint object with \texttt{must\_go=[backup\_sw]} but without explicit \texttt{src}/\texttt{dest}. The validator then fails a check that requires at least one concrete flow target; it detects a \emph{no-op} network policy (no applicable flows) and reports the intent as not enforced.
  \end{quote}

  \item \textbf{Hallucinated identifiers:} Some failures stem from invented labels or policy names (e.g., \texttt{eu\_region}) not present in the cluster.
  
  \begin{quote}\footnotesize
  \emph{Example:} For an intent ``keep PHI in the EU'', the model outputs \texttt{nodeSelector: \{region: eu\_region\}}. The validator cross-checks all referenced labels against the live node-label inventory and fails the check because \texttt{eu\_region} does not exist on any node, implying the placement constraint cannot be satisfied.
  \end{quote}
  
  \item \textbf{Partial topology awareness:} Location-scoped constraints can be under-specified. For example, excluding a named region might miss intermediary transit devices that are labeled differently (e.g., a path traverses a ``region-b'' aggregation switch even though the user intended to exclude the entire country/area). These failures are not due to path-search optimality, but due to incomplete or inconsistent labeling assumptions; they highlight the importance of an explicit label schema and validator-side cross-checks over realized paths. In practice, we found two robust mitigations: (i) enumerate all relevant locations/switches explicitly (or expand a geographic intent into an explicit exclusion set), and (ii) strengthen the label-resolution step so that all devices are consistently annotated before the LLM and validator are run.
\end{enumerate}

Re-running failed intents with either sentence decomposition or explicit enumeration produced passing results, confirming that disambiguation and stepwise prompting are effective counter-measures.

\section{Conclusions and Future Work}

This paper presented a novel framework for privacy-aware intent orchestration using large language models. Our contributions include: (1) an LLM-driven framework for intent-based privacy preservation across both compute and network layers, (2) coordinated scheduling ensuring end-to-end compliance, (3) a prototype implementation with automated validation, and (4) a benchmark dataset of 90 privacy intents.
The convergence of AI and privacy engineering showcased here represents an exciting frontier. Our ultimate vision is a world where users' privacy preferences are upheld not by manual configuration of complex policies, but by simply stating intents and having AI reliably translate them into enforcement---a synergy of human intent and machine execution that keeps technology aligned with our fundamental need for privacy.
The key findings demonstrate that GPT-4o can translate natural-language privacy intents into enforceable Kubernetes and ONOS configurations with 95.6\% accuracy while maintaining interactive latency of approximately 21 seconds. 

Our evaluation uses a controlled testbed and researcher-crafted intents, so results may not fully generalize to production-scale, multi-tenant environments. The approach also relies on label-schema portability across clouds and on the integrity of controller-visible state; when using externally hosted LLMs, operators must consider data minimization/redaction and API cost. Finally, a 95.6\% success rate is promising but not sufficient for fully autonomous configuration in high-criticality settings; deployments may require fail-closed validation, staged rollout, and human approval for critical intents.

We plan to explore automatic intent decomposition and domain adaptation/fine-tuning to reduce downstream parsing and grounding errors, evaluate privacy-protection effectiveness and comparisons against manual configuration baselines, and validate the approach in domain settings such as healthcare IoT, regulated finance, and smart manufacturing.

\bibliographystyle{ACM-Reference-Format}
\bibliography{references}

\end{document}